\documentclass[runningheads]{llncs}
\usepackage[T1]{fontenc}
\usepackage{graphicx}
\usepackage{booktabs}
\usepackage{multirow}
\begin{document}
%
\title{On \emph{de novo} Bridging Paired-end RNA-seq Data}
%
%
\author{Xiang Li\inst{1} \and Mingfu Shao\inst{1,2}}
%
%
\institute{Department of Computer Science and Engineering, The Pennsylvania State University, University Park, PA 16802, USA\\
\and Huck Institutes of the Life Sciences, The Pennsylvania State University, University Park, PA 16802, USA \\
\email{\{xpl5168, mxs2589\}@psu.edu} }

\maketitle              
\begin{abstract}
The high-throughput short-reads RNA-seq protocols often produce paired-end reads, 
with the middle portion of the fragments being unsequenced.
We explore if the full-length fragments can be computationally reconstructed from the sequenced two ends
in the absence of the reference genome---a problem here we refer to as \emph{de novo} bridging. 
Solving this problem provides longer, more informative RNA-seq reads, and benefits downstream 
RNA-seq analysis such as transcript assembly, expression quantification, and splicing differential analysis. 
However, \emph{de novo} bridging is a challenging and complicated task owing to alternative splicing, transcript noises, 
and sequencing errors. It remains unclear if the data provides sufficient information for accurate bridging, 
let alone efficient algorithms that determine the true bridges.
Methods have been proposed to bridge paired-end reads in the presence of reference genome~(called reference-based bridging),
but the algorithms are far away from scaling for \emph{de novo} bridging
as the underlying compacted de Bruijn graph~(cdBG) used in the latter task often contains millions of vertices and edges.
We designed a new truncated Dijkstra’s algorithm for this problem, 
and proposed a novel algorithm that reuses the shortest path tree to avoid running the 
truncated Dijkstra’s algorithm from scratch for all vertices for further speeding up. 
These innovative techniques result in scalable algorithms that can bridge all paired-end reads 
in a cdBG with millions of vertices.
Our experiments showed that paired-end RNA-seq reads can be accurately bridged to a large extent. 
The resulting tool is freely available at https://github.com/Shao-Group/rnabridge-denovo.

\end{abstract}
\section{Introduction}
The high-throughput RNA sequencing technologies~(RNA-seq) enable
accurate measurement of isoform-level gene activities and have been widely used
in biological and biomedical research.
The second generation short-reads RNA-seq,
which remains \emph{de facto} standards for most expression studies,
often produces {paired-end} reads. Such data reports sequences of the two ends of a fragment of an RNA molecule, but misses
the middle portion of the fragment.
The fact that two ends are from the same molecule and that the length of fragments follows
a certain distribution~(through fragment size selection)
provides valuable long-range information in determining complicated splicing variants,
and has been incorporated into various RNA-seq analysis tools and software to improve accuracy,
including splicing-aware alignment~(e.g., STAR~\cite{star}, HISAT2~\cite{hisat}, SpliceMap~\cite{splicemap}),
expression quantification~(e.g., Salmon~\cite{salmon}, kallisto~\cite{kallisto}, RSEM~\cite{rsem}),
transcript assembly~(e.g., StringTie~\cite{stringtie}, StringTie2~\cite{stringtie2}, TransComb~\cite{transcomb}, Scallop~\cite{scallop},
and Scallop2~\cite{scallop2}),
gene fusion detection~(e.g., FuSeq~\cite{fuseq}, STAR-Fusion~\cite{starfusion}, SQUID~\cite{squid}), and 
splicing quantification~(e.g., DARTS~\cite{darts}, leafCutter~\cite{leafcutter}),
among many others. 

We explore computationally inferring the full-length fragments
from paired-end RNA-seq reads, a problem we refer to as \emph{bridging}.
Solving this problem can substantially benefit downstream RNA-seq analysis
such as transcript assembly, isoform quantification, and splicing quantification.
Specifically, the inferred full-length fragments likely contain more splicing junctions than
individual reads, and hence provide additional long-range information that helps resolve more complicated
splicing variants in transcript assembly. 
Longer sequences will be less likely to be ambiguously located to transcripts expressed from the same gene or from homologous genes,
and hence improves isoform quantification.
The reconstructed full-length fragments may reveal missing junctions in the unsequenced portion,
which will likely lead to a more accurate estimation of junction abundance, and hence improves splicing quantification.

The above bridging problem has been studied in a reference-based setting, implemented
as part of the Scallop2 assembler~\cite{scallop2}. In that method, 
the reads alignments of a gene locus
are first organized by a \emph{splice graph}, and reads are
then presented as paths in the graph.
Scallop2 proposed a formulation that seeks a path connecting the two ends of a paired-end read
such that the \emph{bottleneck} weight~(defined as the smallest edge-weight)
of the path is maximized. Scallop2 designed a dynamic programming algorithm, 
and it was demonstrated to be efficient in improving
the accuracy of reference-based assembly.

In this work, we explore the bridging problem in the \emph{de novo} setting, i.e.,
without using a reference genome, termed \emph{de novo} bridging. 
This is motivated by the RNA-seq analysis
for non-model species, for which a good-quality reference genome is not yet available.
The \emph{de novo} bridging problem can also be modeled as a graph problem.
Specifically, all RNA-seq reads can be organized by
a {de Bruijn} graph~(dBG) or compacted {de Bruijn} graph~(cdBG)~\cite{bcalm2}.
(In this paper we use cdBG.)
Similar to the aforementioned reference-based bridging,
the sequenced two ends of a fragment can be mapped to the cdBG,
and the bridging problem amounts to finding a path that can connect the two ends in the graph.
Given the proven efficiency of the formulation proposed in Scallop2 for {reference-based} bridging,
here we adopt it for \emph{de novo} bridging, i.e., to seek a connecting path
in the cdBG such that the bottleneck weight is maximized. However, the dynamic programming
algorithm proposed in Scallop2 cannot scale for \emph{de novo} bridging,
as the underlying cdBG graph may contain millions of vertices~(while
splice graphs typically contain hundreds of vertices or less).

We propose two algorithmic innovations to enable \emph{de novo} bridging
that scales to graphs with millions of vertices.
First, we design a \emph{truncated Dijkstra's algorithm} which can find the optimal
path while taking into account the fragment length to speed up searching.
Second, we propose to use shortest path tree to store the optimal solutions for a single vertex
and propose an efficient algorithm to construct such shortest path tree 
for the next vertex by reusing the nodes and edges on previous trees. 
This allows us to avoid running the above
truncated Dijkstra’s algorithm from scratch for all possible vertices.
Combined, the resulting algorithm can accurately bridge all paired-end reads
in a couple of hours on typical RNA-seq samples~(with millions of vertices
in the underlying cdBG).

\section{Algorithms} \label{sec:algorithm}

\begin{figure}[!b]%
\centering{\includegraphics[width=1\textwidth]{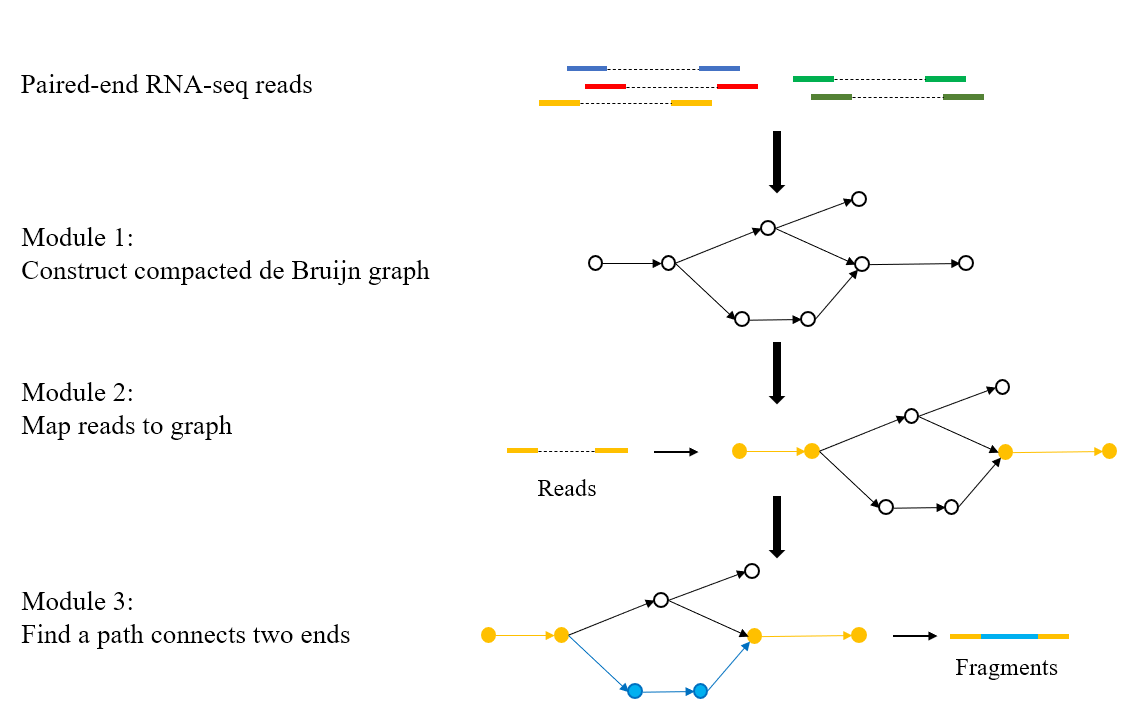}}%
\caption{The pipeline of our method for \emph{de novo} bridging.}%
\label{fig:pipe}%
\end{figure}

Our approach for \emph{de novo} bridging
consists of three modules~(see Figure~\ref{fig:pipe}): constructing the cdBG~(Section~\ref{subsec:constructing}), 
mapping reads to the cdBG~(Section~\ref{subsec:mapping}), and bridging the reads.
The last module is organized as a formulation~(Section~\ref{subsec:formulation})
followed by the algorithms for solving the formulation~(Section~\ref{subsec:bridging}).

\subsection{Constructing Compacted de Bruijn Graph~(cdBG)} \label{subsec:constructing}

\begin{figure}[h]%
\centering{\includegraphics[width=0.6\textwidth]{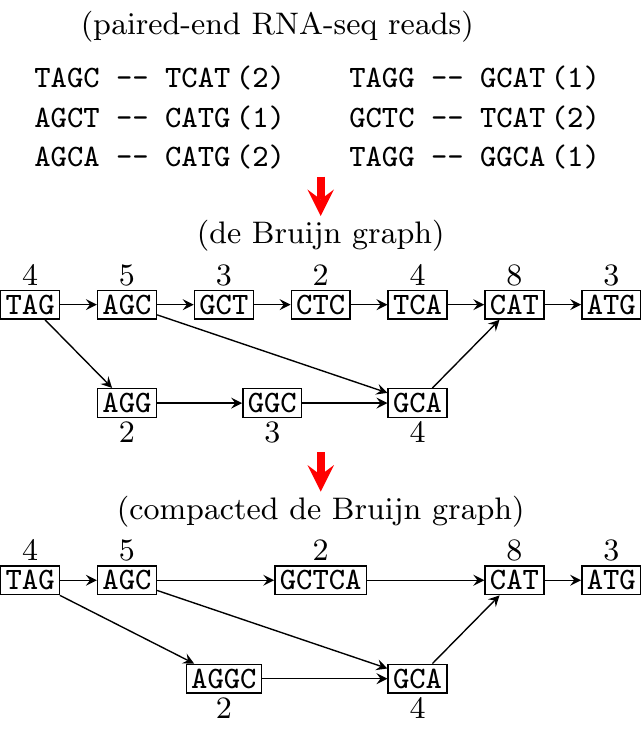}}%
\caption{Example of a cdBG constructed from paired-end RNA-seq reads.}%
\label{fig:cdbg}%
\end{figure}

We use cdBG to represent the given paired-end RNA-seq reads.
See Figure~\ref{fig:cdbg}. In the {de Bruijn} graph~(dBG), each vertex represents a distinct $k$-mer,
and its weight is equal to the number of the appearance in the reads. 
The corresponding cdBG is defined as
concatenating each simple path~(i.e., every vertex
in it except the first and the last one has in-degree of 1 and out-degree of 1) of the dBG
as a single vertex~(the resulting sequence is called a \emph{unitig}). 
To comply with our formulation of finding the most reliable path~(see details in Section~\ref{subsec:formulation}),
we assign the weight of each vertex in cdBG as the {smallest} weight of the corresponding simple path in dBG.

In implementation, we directly construct the cdBG and calculate
vertex weights, instead of explicitly constructing dBG as an intermediate.
Specifically, we use the library Bifrost~\cite{bifrost} to build the cdBG.
In order to assign vertex weights, we first build
a table that stores the frequency of each $k$-mer;
we then examine all $k$-mers in this vertex~(a unitig of length $l$ contains $l-k+1$ $k$-mers) 
and assign the smallest frequency as the weight of the vertex.

\subsection{Mapping Reads and Constructing Equivalent Classes} \label{subsec:mapping}
After constructing the cdBG, we map all paired-end reads to the
graph. For each end, we first call the findunitig function provided in the library Bifrost 
to fetch the vertex that matches the beginning of this end. 
Then we can traverse the graph by extending the following characters to get a list of vertices that this end is matched to.
This procedure gives the path for each end in the cdBG.

Let $G = (V,E)$ be the cdBG and let $f$ be a fragment.
Each end of $f$ can now be represented as a path in $G$, and $f$ can then be
represented as a pair of paths in $G$.  Note that multiple fragments may
correspond to the same pair of paths in $G$; we cluster them into
\emph{equivalent classes}.  In other words, an equivalent class is a pair of paths
$(p_1, p_2)$ in $G$ that represent all fragments with two ends corresponding to
$p_1$ and $p_2$ respectively. 

Let $F = (p_1 = (a_1, a_2,\cdots, a_m), p_2 = (b_1, b_2, \cdots, b_{n}))$ be an equivalent class.
The problem of bridging fragments in $F$ becomes finding a path
in $G$ from $a_m$ to $b_1$; such path is defined as a \emph{bridging path}.
We assume that all fragments in an equivalent class have the same true bridging path.
This is because fragments in an equivalent class are
similar, as their two ends are mapped to exactly the same list of vertices in the graph.
Algorithmically, this assumption allows us
to reduce computational load, as all fragments
can be bridged in a single run.

\subsection{Formulation} \label{subsec:formulation}


We now formulate the \emph{de novo}
bridging as an optimization problem, 
using the same idea proposed in Scallop2~\cite{scallop2}.
We define a full ordering of all bridging paths w.r.t.\ an equivalent class
$F = (p_1 = (a_1, a_2,\cdots, a_m), p_2 = (b_1, b_2, \cdots, b_{n}))$.
Let $q_1$ and $q_2$ be two arbitrary paths from $a_m$ to $b_1$ in graph $G$.
Let $w_1^i$~(resp.\ $w_2^i$) be the $i$th smallest weight 
in path $q_1$~(resp.\ $q_2$).
We say $q_1$ is \emph{more reliable} than $q_2$, if there exists an integer $k$
such that $w_1^i = w_2^i$ for all $1\le i < k$, and $w_1^k > w_2^k$.
We now formulate the bridging problem as to find the most reliable path.
Intuitively, we seek a path $q$ from $a_m$ to $b_1$ in $G$ such that the smallest weight in this path
is maximized, and in case there are multiple
paths with maximized smallest weight, among them we seek the one whose second smallest weight is
maximized, and so on. This formulation has been showed to be accurate 
when applied for reference-based bridging.


We note that this formulation satisfies the \emph{optimal substructure}
property. Specifically, if $a_m \to v_{i_1} \to v_{i_2} \to \cdots \to v_{i_k}
\to b_1$ is the most reliable path, then $a_m\to v_{i_1} \to v_{i_2}\to \cdots
\to v_{i_k}$ is the most reliable path from $a_m$ to $v_{i_k}$.  This formulation
forms the basis for designing efficient algorithms.
The bottleneck weight~(smallest weight of the optimal path) in this formulation
can be used as a filtering criterion to decide if a bridging path is true.  We
set this threshold with different numbers to balance sensitivity and precision
in bridging~(see Table~\ref{tab:simu2} and Table~\ref{tab:real2}).

 
\subsection{Bridging Algorithms} \label{subsec:bridging}
Given the \emph{optimal substructure} property, a straightforward
dynamic programming can be designed~(which essentially was used in Scallop2).
However, as the cdBG constructed from a typical RNA-seq sample
may consist of millions of vertices~(see Table~\ref{tab:time2}), above dynamic programming algorithm simply cannot scale.
We propose to adopt the \emph{Dijkstra's algorithm}.
Dijkstra's algorithm is primarily used for solving the shortest path problem, 
but we modify it to find the most reliable path defined above. 
Specifically,  we maintain an array $d[]$ to store the bottleneck weight on the most
reliable path from the source to other vertices;
each time we select the vertex with highest $d$-value using a priority queue.
Then for each vertex $j$ adjacent to $i$, $d[j]$ can be updated as $d[j] =
\max(d[j], \min(d[i], e(i,j)))$ where $e(i,j)$ is the weight of the edge from
vertex $i$ to vertex $j$.  

A single run of the above {Dijkstra's algorithm} starting from a vertex
in the cdBG $G = (V, E)$
will find the most reliable path connecting it to all other vertices.  It runs
in $O(\left|V\right|\log{\left|V\right|})$ time, already faster than the dynamic
programming algorithm used in Scallop2, which runs in $O(|V'|\cdot|E'|)$ time where
$V'$ and $E'$ are the vertices and edges of the splice graph.

To further speed it up, we propose two algorithmic innovations.
First, we implemented a \emph{truncated Dijkstra's algorithm} to find the most reliable bridging
path starting from a starting vertex $a_m$ to any other vertex up to a certain length $D$~(default value is $400 + 2L$ in our implementation, where $L$ is read length).
This is to incorporate the prior knowledge that fragment length follows an empirical distribution
and an upper bound of fragment length can be assumed. In our truncated Dijkstra's algorithm,
for each vertex $v$ we maintain the total length of the most reliable path from the current starting $a_m$ to $v$,
and when such length for $v$ reaches $D$, 
we won't further extend from $v$ to any other vertices in the Dijkstra's algorithm.

\begin{figure}[b]%
\centering{\includegraphics[width=0.65\textwidth]{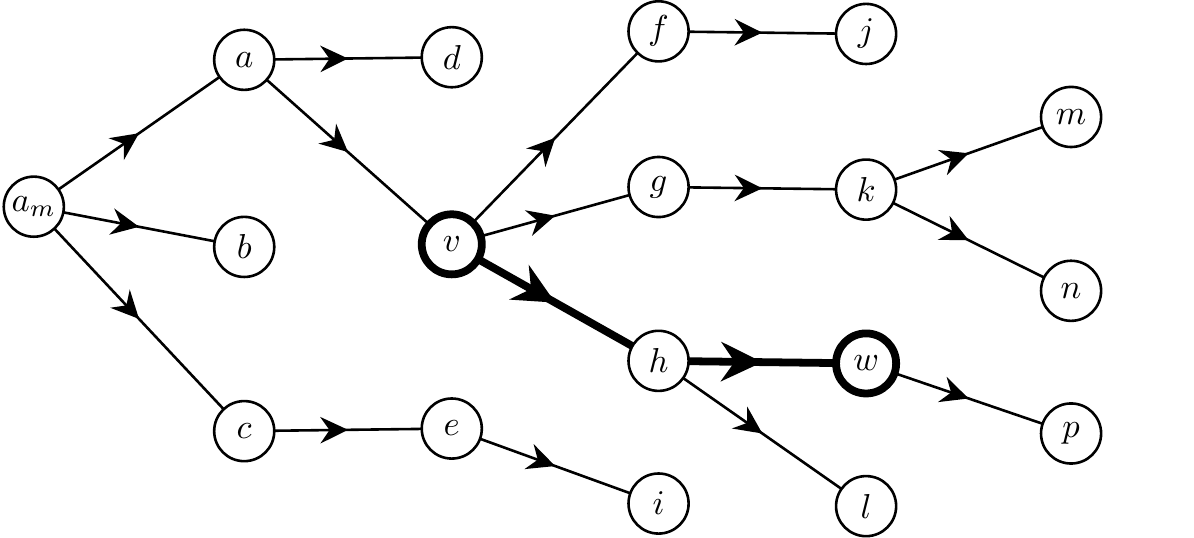}}%
\caption{Illustrating the shortest path tree starting from vertex $a_m$.
Consider subtree rooted at $v$. Then for any vertex in it, say $w$, the path
from $v$ to $w$ in this subtree is the most reliable path from $v$ to $w$
in the cdBG.}%
\label{fig:subtree}%
\end{figure}

The second algorithmic innovation is that we reuse the optimal solutions obtained for the current starting vertex $a_m$
to construct the optimal solutions for the next starting vertex $a_m'$. This allows us to avoid running the
above truncated Dijkstra's algorithm from scratch for all possible starting vertices. Specifically, for
the current starting vertex $a_m$, we maintain a \emph{shortest path tree}, denoted as $T(a_m)$, to store 
the most reliable paths from $a_m$ to all other vertices~(up to length $D$), i.e.,
the unique path in $T$ from root $a_m$ to any vertex $v$ gives the most reliable path
from $a_m$ to $v$ in the cdBG.  See Figure~\ref{fig:subtree}. 
This tree can be constructed in linear time while running the truncated Dijkstra's algorithm.
We note that, again according to the property of optimal substructures, any subtree of $T(a_m)$, say the one rooted
at $u$, also gives the most reliable path from $u$ to any other vertex $w$ in the subtree.
This suggests we can reuse the subtree rooted at $u$ to calculate all reliable paths starting from $u$~(and then construct the corresponding shortest path tree).
More specifically, in implementation, we directly load the subtree rooted
at $u$ to the priority queue~(recall that the core data structure of Dijkstra's algorithm is a
priority queue that gets updated iteratively). In other words, for the next starting vertex $u$, we
run the truncated Dijkstra's algorithm in the middle rather than from scratch, as we already know
the optimal solutions for a subset of vertices~(i.e., those in the subtree of $T(a_m)$ rooted at $u$).
To benefit from this property to the largest extent, 
we determine the starting vertex whose subtree is largest as the next one to bridge.

\section{Results} \label{sec:result}

\subsection{Resulting Tool} \label{subsec:tools}

The above algorithm was implemented, available at 
https://github.com/Shao-Group/rnabridge-denovo.
The input files for this tool is the paired-end RNA-seq data in fastq/fasta format, and it
generates sequences of full fragments again in fasta format. 
Since so far our method is the only one for \emph{de novo} bridging,
we could not compare it with other methods in the following experiments.

\subsection{Datasets} \label{subsec:datasets}

We use two datasets to evaluate the accuracy of bridging. 
The first dataset includes 80 paired-end RNA-seq samples simulated using Flux-Simulator~\cite{fluxsimulator}.
We vary two parameters in the simulation: the average length of fragments~(flen; 300 and 500)
and the read length~(rlen; 75 and 100). 
For each combination of parameters, we independently simulated
20 samples. The number of reads in samples with fragment length being 300 and 500 are roughly 150M and 90M, respectively.
The second dataset was previously used in the Scallop paper~\cite{scallop}:
it contains 10 biological RNA-seq samples.

\subsection{Results on Simulation Data} \label{subsec:result}

We first evaluate the accuracy of our algorithm using simulation data,  
for which the ground-truth is available. 
A bridged fragment is correct only if it is exactly the same as the ground-truth fragment. 
The sensitivity is defined as the number of correctly bridged fragments divided by the total number of fragments~(i.e., paired-end reads),
and precision is defined as the number of correctly bridged fragments divided by the total number of bridged fragments.

The results are summarized in Table~\ref{tab:simu2}.
The bottleneck threshold used to filter bridges is set to 5 for all simulated samples.
Our algorithm exhibits high accuracy, suggesting that the missing portion can be accurately
bridged given solely the reads information and that our algorithm is able to use such information
to bridge.  The accuracy drops with long fragment length. This is expected
as in this case the missing portion is longer and therefore harder to bridge. 
Higher accuracy is observed when the read length
is longer at the same fragment length, again as expected.

\begin{table}[h]
\def\arraystretch{1.2}%
\setlength\tabcolsep{5pt}%
\caption{Averaged bridging accuracy on the simulated RNA-seq datasets.} \label{tab:simu2} 
\centering \begin{tabular}{@{}rrccc@{}}
\toprule 
flen & rlen & bottleneck & sensitivity~(\%) & precision~(\%)\\
\midrule
300 & 75  & 5 & 80.7 & 91.8 \\
300 & 100 & 5 & 85.7 & 92.5 \\
500 & 75  & 5 & 66.6 & 85.4 \\
500 & 100 & 5 & 76.3 & 86.9 \\
\bottomrule
\end{tabular}
\end{table}

\subsection{Results on Real Data} \label{subsec:result2}

We then evaluate the accuracy of our algorithm on real dataset.
As we do not have ground-truth for them, we use the sequences in the reference transcriptome
to evaluate.  We align all the bridged fragments to reference using BLAT~\cite{kent2002blat}. 
A bridged fragment is correct only if it is hit by one of the reference sequences with at least
95\% sequence identity. The sensitivity and precision are defined the same as in evaluating with simulation data.

\begin{table}[!b]
\def\arraystretch{1.2}%
\setlength\tabcolsep{5pt}%
\caption{The bridging accuracy on the 10 RNA-seq samples.
The number of paired-end reads are given in unit of million.} \label{tab:real2} 
\centering \begin{tabular}{@{}rrrcc@{}}
\toprule 
SRA ID & \#paired-reads & bottleneck & sensitivity~(\%) & precision~(\%) \\
\midrule
SRR307903 &	36.0M & 20 &	65.3 & 91.8 \\
SRR307911 &	41.4M & 20 &	42.7 & 91.3 \\
SRR315323 &	30.3M & 20 &	27.9 & 87.9 \\
SRR315334 &	39.5M & 20 &	49.5 & 93.3 \\
SRR545723 &	38.9M & 20 &	42.7 & 81.6 \\
SRR387661 &	124M  & 100&	58.2 & 90.1 \\
SRR534291 &	114M  & 100&	64.3 & 93.5 \\
SRR534307 &	165M  & 100&	50.6 & 74.7\\
SRR534319 &	76.6M & 100 &	24.6 & 69.5 \\
SRR545695 &	119M  & 100&	37.6 & 75.5 \\
\bottomrule
\end{tabular}
\end{table}

The results are summarized in Table~\ref{tab:real2}.
Overall the precision keeps high but the sensitivity varies quite a lot.
This is because we prioritize precision by setting a high bottleneck-threshold: 
20 for samples with less than 50M paired-end reads and 100 otherwise.
Comparing with simulated data, biological data are noisier and harder to bridge.
We note that high precision is more desirable for downstream analysis in order not to 
introduce false positives. Given the high precision, even a small portion of correctly bridged paired-end
reads are able to improve the accuracy of downstream applications.
Users may also choose to adjust the bottleneck-threshold to balance the precision and sensitivity.

\subsection{Analysis of Running Time} \label{subsec:result3}

As mentioned in Section~\ref{sec:algorithm}, our tool consists of 3 modules.
We show the breakdown of the CPU time of the 3 modules together
with the size of the resulting cdBGs on the 10 biological samples in Table~\ref{tab:time2}.
Note that module~3 takes the least CPU time among 3 modules,
which proves the efficiency and scalability of the bridging algorithms.

\begin{table}[h]
\def\arraystretch{1.2}%
\setlength\tabcolsep{8pt}%
\caption{The size~(number of vertices and edges; in unit of million) of the 
cdBGs and the CPU time~(in minutes)
measured for the 3 modules of our bridging tool on the 10 biological RNA-seq samples.} \label{tab:time2} \centering 
\begin{tabular}{@{}crrrrrr@{}}
\toprule 
\multirow{2}{*}{SRA ID} & \multicolumn{2}{c}{size of cdBG} & \multicolumn{3}{c}{CPU time~(in minutes)} \\
& \#vertices & \#edges & module~1 & module~2 & module~3\\
\midrule
SRR307903 &	5.15M &	6.99M &	149 &	57 &	26 \\
SRR307911 &	8.11M &	10.71M &	175 &	66 &	75 \\
SRR315323 &	7.68M &	9.45M &	120 &	40 &	63 \\
SRR315334 &	6.35M &	10.16M &	158 &	67 &	37 \\
SRR387661 &	18.71M & 27.36M &	515 &	223 &	141 \\
SRR534291 &	15.14M &	26.67M &	685 &	370 &	128 \\
SRR534307 &	34.29M &	56.00M &	1038 &	561 &	281 \\ 
SRR534319 &	16.48M & 22.09M &	311 &	118	 & 121 \\
SRR545695 &	19.43M & 27.49M &	479 &	201 &	174 \\
SRR545723 &	13.47M & 16.04M & 243 &	99 &	159 \\
\bottomrule
\end{tabular}
\end{table}

\section{Conclusions and Discussion} \label{sec:conclusions}

In this work, we study the problem of reconstructing the
full-length fragments from paired-end reads without a reference. The experimental results
showed that the simulated RNA-seq data does provide
sufficient information for accurate bridging. As for the real RNA-seq data, it
provides enough information to bridge a large part of the reads. 

We invented two algorithmic innovations for \emph{de novo} bridging,
including a novel truncated Dijkstra’s algorithm and a new technique 
that reuses optimal path trees to speed up.
Experimental results proved that the resulting tool is efficient on real data.
We also proved that the formulation used for reference-based bridging 
is also accurate for \emph{de novo} bridging when applied on the cdBG;
this conclusion was unclear before we conducted this
research. 

We explored if bridging could improve
\emph{de novo} transcript assembly. To this end, we
piped the bridged fragments to one leading assembler TransLiG~\cite{translig},
but only observed marginal improvement. This may be because
TransLiG is not optimized to make use of mixed short and long sequences.
Developing a new \emph{de novo} assembler that can fully use such bridged
data is on our research agenda. Experimenting if \emph{de novo} bridging
could improve isoform quantification and splicing quantification
is also an interesting future research topic.

The sensitivity of our algorithm is low on some biological RNA-seq samples.
One reason is that we used a high bottleneck-threshold to keep high precision,
which consequently disconnects paired-end reads in low-coverage gene loci.
We are therefore developing a post-bridging algorithm to use full-range information
in the reads to decide if a bridged fragment is correct~(rather than
just using a bottleneck-threshold), in the hope of keeping high precision while improving sensitivity.
Specifically, note that the cdBG is not a loss-free representation of sequencing reads,
as it breaks reads into $k$-mers, and any phasing information
beyond $(k+1)$-mer is not represented.  Let $s$ be the bridged
sequence of a fragment $f$ constructed using above algorithm.
We can examine each sliding window of length $L$, where $L$ is the read length,
and determine the number of input reads that are identical to this $L$-mer~(i.e., 
these reads \emph{support} this $L$-mer). This gives a \emph{supporting profile},
a vector of length $|s| - L + 1$ for this bridged sequence.
Intuitively, a profile with few 0s suggests that the bridged sequence is likely a
true one, while long consecutive 0s in the profile suggest a false bridge.
We are experimenting if such a supporting profile could lead to more efficient
algorithms in boosting bridging accuracy.

\section*{Acknowledgments}

This work is supported by the US National Science Foundation (2019797 and 2145171 to M.S.)
and by the US National Institutes of Health (R01HG011065 to M.S.).

%
%
%
\bibliographystyle{splncs04}
\bibliography{compbio}

\end{document}